\documentclass[12pt,preprint,longnamesfirst]{aastex}
\usepackage{pslatex}
\usepackage[latin1]{inputenc} 

\shorttitle{The Decay of Accreting Triples}
\shortauthors{Umbreit et al.}

\begin{document}

\title{The Decay of Accreting Triple Systems as Brown Dwarf Formation Scenario}

\author{Stefan Umbreit}

\affil{Max-Planck-Institut für Astronomie, Königstuhl 17, D-69117
Heidelberg, Germany}

\email{umbreit@mpia.de}

\author{Andreas Burkert}

\affil{Universitäts-Sternwarte München, Scheinerstrasse
1, D-81679 München, Germany}

\email{burkert@usm.uni-muenchen.de}

\author{Thomas Henning}

\affil{Max-Planck-Institut für Astronomie, Königstuhl 17, D-69117
Heidelberg, Germany}

\email{henning@mpia.de}

\author{Seppo Mikkola}

\affil{Tuorla Observatory, University of Turku, 21500 Piikkiö , Finland}

\email{seppo.mikkola@astro.utu.fi}

\and{}

\author{Rainer Spurzem}

\affil{Astronomisches Rechen-Institut Heidelberg, Mönchhofstraße 12-14, D-69120 Heidelberg,
Germany}

\email{spurzem@ari.uni-heidelberg.de}

\begin{abstract} We investigate the dynamical decay of non-hierarchical
accreting triple systems and its implications on the ejection model as
Brown Dwarf formation scenario. A modified chain-regularization scheme
is used to integrate the equations of motion, that also allows for
mass changes over time as well as for momentum transfer from the
accreted gas mass onto the bodies. We integrate an ensemble of triple
systems within a certain volume with different accretion rates,
assuming several prescriptions of how momentum is transferred onto the
bodies. We follow their evolution until the systems have
decayed. We analyze the end states and decay times of these systems
and determine the fraction of Brown Dwarfs formed, their escape speeds
as well as the semi-major axis distribution of the formed Brown Dwarf
binaries. We find that the formation probability of Brown Dwarfs
depends strongly on the assumed momentum transfer which is related to
the motion of the gas. Due to ongoing accretion and consequent
shrinkage of the systems, the median escape velocity is increased by a
factor of 2 and the binary separations are decreased by a factor of 5
compared with non-accreting systems. Furthermore, the obtained
semi-major axis distribution drops off sharply to either side of the
median, which is also supported by observations. We conclude that
accretion and momentum transfer of accreted gas during the dynamical
decay of triple systems is able to produce the observed distribution
of close binary Brown Dwarfs, making the ejection model a viable
option as Brown Dwarf formation scenario. \end{abstract}


\section{\label{sec:Introduction}Introduction}

In the past few years many Brown Dwarfs have been detected (e.g. \citealp{2000ARA&A..38..485B}).
Brown Dwarfs, known as 'failed stars' which have not enough mass to
start hydrogen burning as in normal stars, were found at many different
star-formation sites like Taurus \citep{2002ApJ...580..317B}, Orion
(e.g. \citealp{2002ApJ...573..366M,1999ApJ...521..671B}), Ophiuchus
(e.g. \citealp{2002ApJ...566..993A}) and the Chamealeon cloud (e.g.
\citealp{2004A&A...416..555L,2000A&A...359..269C}), as cluster members
\citep{2002sf2a.conf..469M,1998A&A...336..490B,1998ApJ...507L..41M}
and as free-floating objects \citep{1999ApJ...519..802K,2000AJ....120..447K}.
Based on the frequency of detection it is widely believed that they
should be as common as low-mass stars. In addition to their similar
abundance many of them also show accretion features like ordinary
TT stars and it was even possible to detect circumstellar disks around
them \citep{2003AJ....126.1515J,2003ApJ...590L.111P,2003ApJ...593L..57K,2002ApJ...573L.115A,2001A&A...376L..22N}.
Also some of them are known to form binary and higher order systems
\citep{2003AJ....126.1526B}.

Based on these results it is tempting to claim that Brown Dwarfs may
have been formed like ordinary TT stars, especially since their accretion
features, i.e. mainly H$\alpha$ emission, vanish after about the
same time as it is the case with TT stars (see e.g. \citet{2003ApJ...585..372L}).
Indeed most of these observed properties of Brown Dwarfs can be understood
this way. The only feature that lacks a clear understanding are the
properties of Brown Dwarf binaries. \citet{2003MNRAS.346..354K} argue
that if Brown Dwarfs are formed like low-mass stars they should have
the same pairing statistics and binary properties, scaled down to
the substellar regime. On the contrary, from recent surveys of \citet{2003ApJ...587..407C},
\citet{2003AJ....125.3302G}, \citet{2003AJ....126.1526B} and
\citet{2003ApJ...594..525M}, as well as from cluster simulations of
\citet{2003MNRAS.346..354K} it has been found that the
observed properties of Brown Dwarf binaries cannot be obtained by
extending the simple pairing rules of the standard star-formation
model into the substellar regime \citep{2003MNRAS.346..354K}. Therefore
\citet{2003MNRAS.346..354K} come to the conclusion that Brown Dwarfs
may not be formed with the same scaled down properties as stars and
further infer that, in order to form Brown Dwarfs, their accretion
phase must be interrupted by other (external) processes. Which process
this actually can be is currently under vivid debate. It has been
suggested that the strong, ionizing UV radiation of hot O and B stars
might be responsible for the increased number of Brown Dwarfs in Orion
\citep{2003MNRAS.346..369K}. For low-mass star forming regions harboring
Brown Dwarfs, like the Chamaeleon cloud, there must be a different
process at work because of the generally low abundance of higher mass
stars and because it is expected that photo evaporation is only efficient
in the vicinity of massive stars \citep{2003MNRAS.346..369K}. It
has been suggested, by many different authors, that substellar objects
can be formed by fragmentation of massive circumstellar disks \citep{1998Sci...281.2025L,2001ApJ...551L.167B,1998MNRAS.300.1189B,1998MNRAS.300.1205W,1998MNRAS.300.1214W,2000ApJ...540L..95P}.
Such a process is also seen in simulations of collapsing molecular
clouds \citep{2002MNRAS.332L..65B,2003PhDT.........1L}. It is currently
the most favorable one to explain the increased number of low-mass
Brown Dwarfs with a mass at around $0.02$M$_{\sun}$ \citep{2002MNRAS.332L..65B,2003PhDT.........1L}
as reported by \citet{2003ARA&A..41...57L}. 

\citet{2001AJ....122..432R} suggested that the ejection of fragments
from unstable multiple systems out of their surrounding molecular
cloud may lead to an early end of the accretion process of the fragments
and, consequently, leave some of them substellar. This formation scenario
is constantly challenged by observational studies \citep[and more]{2002ApJ...580..317B,2001A&A...376L..22N}.
They argue that because accretion features are observed around objects
with an age of up to 10~Myr, which is about the lifetime of disks
around T Tauri stars, close collisions, required for the ejection
of fragments, cannot have happened as they tend to truncate the disks,
severely limiting their lifetime, which in turn should make the frequency
of detection much lower than actually observed. On the other hand
the amount of material that is stripped off the disk is also sensitive
to the excentricity of the perturber orbit. For instance strongly
hyperbolic encounters are known to be much less destructive than parabolic
ones which are assumed to be the most likely encounter orbit in larger
clusters (\citet{1994ApJ...424..292O} and \citet{1990ppfs.work..389L}).
For accreting small-N clusters the encounter parameters are not well
known and certainly need to be studied in greater detail. Based on
these uncertainties and the fact that not all young Brown Dwarfs possess
disks, we believe that the ejection scenario cannot be ruled out completely
but needs more accurate modeling in order to provide better predictions.

The aim of our study is to explore under which conditions Brown Dwarfs
form in triple systems and derive statistics of escaping Brown Dwarfs
and binaries. For this we further investigate the ejection scenario
by means of N-body calculations where the bodies are gaining mass
according to a given accretion rate. Proceeding this way, we do not
address the problem of disks around Brown Dwarfs. In section \ref{sec:Previous-Work}
we briefly highlight some interesting theoretical studies that have
been recently done regarding decaying multiples and their relation
to Brown Dwarf formation as well as other Brown Dwarf formation scenarios.
In section \ref{sec:Analytical-Predictions} we show what we can expect
from the results of the evolution of accreting multiple systems, in
section \ref{sec:Simulations} we explain the initial setup as well
as the methodology of our simulations, in section \ref{sec:Results}
we present and discuss our results and, finally, in section \ref{sec:Conclusions}
we draw conclusions about the ejection scenario as a viable Brown
Dwarf formation scenario.

\section{\label{sec:Previous-Work}Previous Numerical Studies on Brown Dwarf
Formation }

\citet{2002MNRAS.332L..65B} were the first who could follow the fragmentation
of a massive cloud, subject to a turbulent velocity field, down to
substellar masses and therefore were able to draw conclusions about
how Brown Dwarfs form. They find that their Brown Dwarfs formed mainly
through instabilities in massive circumstellar disks and, to a lower
amount, as ejected embryos from unstable small-N clusters, confirming
these Brown Dwarf formation channels. \citet{2003PhDT.........1L}
carried out a similar study but instead of focusing on an entire turbulent
molecular cloud he studied many realizations of a collapsing cloud
core and its fragmentation. In both studies the number of Brown Dwarfs
did not exceed 100 and there were only a few Brown Dwarf binaries.
Their low frequency of less than $10\%$ is in contradiction with
the observed one of $\approx20\%$ by \citet{2003ApJ...587..407C}
and \citet{2003AJ....126.1526B}. The reason for this is not very
clear yet. \citet{2003MNRAS.339..577B} argue that the disruption
of wider Brown Dwarf binaries due to the closeness of the encounters
in their simulation ($<20{\rm AU}$) and exchange interactions with
stars, which replace the lower-mass substellar members with more massive
stellar objects, reduce significantly the number of Brown Dwarf binaries.
However, we cannot exclude the possibility that the softening of the
gravitational potential of the fragments, which is of the order of
$10{\rm AU}$ in their simulation, limits the formation of close Brown
Dwarf binaries with semi-major axis $\leq10{\rm AU}$, due to the
lower absolute value of the potential energy at those radii compared
to the non-softened potential, which should result in wider binary
pairs. The greater semi-major axis of those binaries also reduces
the probability that they survive subsequent gravitational interactions
with the cluster. Given the low absolute numbers of Brown Dwarf binaries
in these simulations it is impossible to obtain any firm statistical
result about Brown Dwarf binaries and to compare them to the observations.

\citet{2003MNRAS.346..369K} discuss various scenarios of Brown Dwarf
formation. They try to explain the different abundances of Brown Dwarfs
seen by \citet{2002ApJ...580..317B} in the Taurus region and the
known one in the Orion nebula cluster \citep{2002ApJ...573..366M,2000ApJ...540.1016L}
by estimating the kinematics of Brown Dwarfs resulting from the different
formation models. They find that the ejection scenario is able to
reproduce the number of Brown Dwarfs per star consistently if one
assumes that in both clusters the same number of Brown Dwarfs are
produced per star and the velocity dispersion of the ejection process
is about 2 km$\cdot$s$^{-1}$. These results have the disadvantage
that they are in disagreement with the estimated Brown Dwarf abundance
in the galactic field, if this abundance is not overestimated \citep{2003MNRAS.346..369K}.
If one, on the other hand, drops the assumption that the Brown Dwarf
production rate is independent of the environment and assumes a velocity
dispersion of $3$km$\cdot$s$^{-1}$ of the ejected Brown Dwarfs,
they find that low-mass tranquil star-forming regions are more efficient
in producing Brown Dwarfs than the ones like the Orion nebula cluster
(ONC). This is also true even if one adds the Brown Dwarfs formed
by photo-evaporation in the ONC. 

The question whether the ejection scenario is able to reproduce the
high abundance as well as the binary properties of Brown Dwarfs has
been addressed by various authors. \citet{2003A&A...400.1031S} calculate
pairing and binary statistics by integrating many small-N clusters
neglecting hydrodynamical interaction by the remaining gas as well
as any ongoing accretion. They constrain their initial conditions
by a modified clump mass spectrum of \citet{1998A&A...336..150M},
which determines the total masses of the clusters, and a composite
single star mass spectrum (SMS) which comprises a recently observed
one for Brown Dwarfs \citep{2001ApJ...556..830B,2002ApJ...567..304C}
as well as one for hydrogen-burning stars \citep{1993MNRAS.262..545K}.
They find broad agreement between their results and observations of
Brown Dwarf binaries, concluding that, once Brown Dwarfs have formed
in sufficient numbers as to fit the observed Brown-Dwarf-IMF of the
galactic field, the subsequent decay of the emerging multiple systems
with the given constraints can explain their binary properties. They
also point out that, because they scale their results by fixing the
virial speed for all systems choosing ${\rm v}_{vir}=3.3$ km$\cdot$s$^{-1}$,
their Brown Dwarf systems are already in a very compact configuration
close to the final binary separations. Indeed for a triple Brown Dwarf
system, all having masses of 0.08M$_{\sun}$ one gets a very small
H$\acute{{\rm e}}$non radius $R_{H}$, which is a measure of the
mean interparticle distance, with a value of $R_{H}\approx10$ AU.
Of course in initially higher order systems the interparticle distances
can be larger, but even for a system of 6 Brown Dwarfs $R_{H}$ is
still as small as 20~AU. This scaling is motivated by the observed
mass versus size relation of molecular cloud cores that imply that
the specific energy and hence the virial speed is a constant and of
about the previously mentioned value. \citet{2003A&A...400.1031S}
simply assume the same relation between the size of the emerging cluster
and its total mass. The findings of \citet{2003A&A...400.1031S} therefore
imply that from a purely dynamical point of view Brown Dwarfs must
have formed in extremely compact configurations in order to explain
the observed Brown Dwarf binary separations of, e.g. \citet{2003ApJ...587..407C}.
It still has to be shown that fragments, which will eventually become
Brown Dwarfs, are initially mostly formed within such small volumes.
Simulations of \citet{2001ApJ...551L.167B} seem to support this view.
\citet{2002MNRAS.336..705B} argue in contrary that at the time the
isothermal collapse of a fragment ends and the gas starts to heat up
and finally halts the collapse, the radii of these fragments should
be at least $5\,{\rm AU}$ and their separations consequently $\gtrsim10\,{\rm AU}$.
They also find in their numerical simulation no binary fragments forming
with a lower initial separation. Also \citet{2002MNRAS.332L..65B}
reported that their ejected Brown Dwarfs were still undergoing significant
accretion before the time of ejection, therefore contradicting the
notion of Brown Dwarfs being {}``frozen'' out of the gas with their
final masses. Furthermore this makes it possible that fragments start
out further apart with lower masses and much lower virial speed, and,
due to their growing masses, finally reach the required compactness
to produce tight binaries. Given these difficulties and the ease at
which close binaries are formed if one includes mass growth during
the dynamical interactions of the fragments, accretion seems to us
the major ingredient to form close Brown Dwarf binaries.

\citet{2003MNRAS.342..926D} focus on accreting multiple system by
placing 5 accreting seeds inside a molecular cloud core following
their evolution in response to the gravitational potential of the
gas as well as the mutual gravitational interaction between the seeds.
They find that Brown Dwarfs, once an appropriate mass function for
the parent cloud cores is chosen, are readily formed if dynamical
interactions with an unstable multiple system are drawn into account.
Even though they came to better statistical results on this formation
scenario by integrating 100 realizations of a multiple system in a
cloud, they got only a few Brown Dwarf binaries. They conclude that
if the binary fraction among Brown Dwarfs turns out to be low, it
can easily be explained by these simulations. On the other hand if
the binary fraction turns out to be high, they infer that the core
mass function must extend down into the substellar regime (the core
mass function of \citet{1998A&A...336..150M} they were using has
a lower cut off of 0.25M$_{\sun}$). Because of their low Brown Dwarf
binary statistics they cannot draw any further conclusions on the
properties of Brown Dwarf binaries. 

Given the computational expense of a full hydrodynamical simulation
and the necessity to include mass growth of the fragments it is certainly
appropriate to treat the gas accretion and interaction in an approximate
fashion by assuming a certain accretion rate and specifying \emph{ad
hoc} the momentum the accreted mass adds to the stellar embryos. This
approach was shown to be a good approximation in \citet{1997MNRAS.285..201B}
if one assumes that the gas is at rest and the bodies are not moving
with significantly supersonic velocities. This modelling of dynamical
interaction of the fragments allows for a better statistical description
of the resulting Brown Dwarf properties, including binaries, because
of the increased number of systems that can be integrated within a
reasonable amount of time. It also quantifies the influence which
accretion has on the dynamical evolution of multiple systems.

\section{\label{sec:Analytical-Predictions}Analytical Predictions}

Our analytical model calculations are mainly based on the toy model
of \citet{2001AJ....122..432R} that we want to discuss here briefly.
These calculations are mainly estimates of timescales of the physical
processes involved. 

At first it is assumed that a flattened cloud is collapsing and the
central region accretes mass at a constant rate of
$\dot{M}_{infall}\sim6\times10^{-6}(T/10{\rm K})^{\frac{3}{2}}{\rm
M_{\sun}{\rm yr^{-1}}}$.  This value was obtained by numerical
simulations of \citet*{1996ApJ...464..387H} who found a period of
nearly constant mass accretion onto the central region that lasted
about 1 free-fall time which corresponds to $\approx10^{5}{\rm yr}$ in
this case. For the following discussion we will consider all physical
processes on this time scale. The central part of the cloud is assumed
to fragment into 3 fragments and the infalling mass is equally
distributed amongst them. The choice of 3 fragments is motivated by
the fact that a triple system is the smallest possible cluster with
the ability to decay into smaller-N systems, and therefore was chosen
as starting point for their investigation. 

Only a few triple systems are known to be stable, while all the other
decay after some time, ejecting a single body and a binary system into
opposite directions.  In order to estimate how many systems have
decayed before their masses grow beyond the substellar limit, which is
set here for simplicity to $0.08 {\rm M_{\sun}}$, one can use that
such a decay can be described like radioactive decay, i.e. the
fraction of systems that have not yet decayed at a time $t$ is a
single exponential function characterized by the half-life time of the
decay \citep{1986Ap&SS.124..217A}. The half-life of such a decay is
directly proportional to the crossing time, which is the typical time
scale of the evolution of a multiple system. It is a function of the
masses of the fragments as well as the total energy of the system,
which by themselves are time dependent quantities. For the total
energy \citet{2001AJ....122..432R} use a simplified expression,
assuming $R$, the mean harmonic distance, to be constant at $200 \rm
AU$. In this model it is further assumed, that the masses of the
fragments grow linearly with time. The resulting decay function is
shown in \citet[Fig.1]{2001AJ....122..432R}.  In this figure it is
also accounted for that, at the beginning of the formation of the
fragments, their interaction potential is, due to their small masses,
still low compared to the potential of the surrounding gas. Therefore
the fragments do not interact significantly until a time $T_{i}$
which, in absence of any detailed hydrodynamical calculations, cannot
be reliably determined and was therefore set to a time when
$M=0.04{\rm M_{\sun}}$.  This value is quite
\char`\"{}pessimistic\char`\"{}, as it significantly reduces the time
interval within which the fragments have to decay before they reach
the hydrogen burning limit. In this calculation a third of the systems
have decayed before they reached that mass limit and consequently
became Brown Dwarfs. At this point we find it is worth pointing out
that lower accretion rates result in a higher abundance of Brown
Dwarfs, although the decay of the triple systems happens a lot later.

This result shows, according to \citet{2001AJ....122..432R}, that the
ejection scenario is able to produce many Brown Dwarfs even under
quite \char`\"{}pessimistic\char`\"{} assumptions and should be
therefore very efficient in more realistic situations. For example,
they note, that the accretion process is stopped earlier if the
infalling gas is not equally distributed among the bodies. A mass
difference between them tends to drive out the lower-mass member by
mass segregation, whereby this body accretes even less material
because of the lower gas densities in the outer regions of the
cloud. This process drives it even further out by mass segregation and
so forth. This scenario has been named 'competitive accretion' and was
investigated by \citet{2001MNRAS.323..785B}.

Here we want to show, however, that the moderate probability of forming
Brown Dwarfs from unstable, equally accreting triple systems is a dramatic
understatement of how efficient this particular model is. The key point
that has been left out in the previous calculation is the fact that while
the fragments are growing in mass the whole multiple system shrinks in
size, making $R$ a function of time. To account for this effect we
calculate the change of energy, making assumptions about the motion of the
accreted gas.

As a first approximation, we consider accretion of gas at rest. 
The resulting equations of motion of the bodies differ, compared to the
equations of motion with constant masses, by the term
$\dot{M}\cdot{\bf v}$ which can also be interpreted as an additional velocity
dependent friction force. In the cluster simulations of
\citet{1997MNRAS.285..201B} this approximation was shown to reflect
the evolution of the total energy of a cluster over time of their
fully hydrodynamical calculation sufficiently well as long as the
bodies do not move significantly supersonic. However, as we expect our
fragments to move with a speed of the order of $1\,{\rm km/s}$, which
is about 4 times higher than the typical sound speed of a cold
molecular cloud core, this approximation still underestimates the
compactness of the systems, as additional drag forces, caused by the
supersonic motion of the bodies, are not taken into account.

For the total energy of a triple system accreting gas at rest we get
\begin{equation}
E(t)=E_{0}\cdot\left(\frac{\dot{M}}{M_{0}}t+1\right)^{5}. 
\end{equation}
with $E_0$ and $M_0$ being the initial value of the total energy and
mass of a fragment, respectively (see Appendix
\ref{sec:Energy-analytic} for a detailed derivation). As is easily seen
the energy depends strongly on the accretion rate $\dot{M}$ and
decreases, due to $E_{0}<0$, with the 5$^{\textrm{th}}$ power of $t$,
making the system shrink extremely quickly. This expression also shows
that the typical time scale at which the systems shrink goes with
$\sim\left(\dot{M}/M_{0}\right)^{5}$, i.e. with the accretion time
scale to the power of 5 for accretion of gas at rest.

If we compare this with the toy model of \citet{2001AJ....122..432R}
where the energy equation reads 
\begin{equation}
E(t)=-3\, G\,\frac{\left(\dot{M}\cdot
t+M_{0}\right)^{2}}{R} 
\label{eq:E(t)RC}
\end{equation}
with constant $R$, it can be clearly seen that we get a less steep
dependence on $t$ and $\dot{M}$ if only the mass growth but not
the time dependence of $R$ is considered, underestimating the compactness
of accreting multiple systems, which in turn makes the decay times
too long.

We also want to see how the energy decreases if the gas is not at
rest. For simplicity we assume that the accreted mass is moving at
the same speed and direction as the bodies, thus leaving the velocities
of them unchanged.  The equations of motion remain
unchanged and the energy equation reads
\begin{equation}
E(t)=E_{0}\cdot\left(\frac{\dot{M}}{M_{0}}\cdot t+1\right)^{3}
\end{equation}
which differs from the solution with accretion of gas at rest only in
the value of the exponent. Now the energy depends less strongly on the
accretion rate and $t$ as the typical time scale at which the system
shrinks goes only with the 3$^{\textrm{rd}}$ power of the accretion
time scale. We should therefore expect, on average, longer decay times
than we would get in the case of accretion of gas at rest,which are,
however, still shorter as in the toy model of
\citet{2001AJ....122..432R}.  For illustration, we can also assume
that the gas has always the same velocity but the opposite direction
as the fragments. This would lead to a value of the exponent of $7$,
resulting in an even stronger decrease of the total energy and
therefore to even shorter decay times as in the
'accretion-of-gas-at-rest' model.

\begin{figure}
\plotone{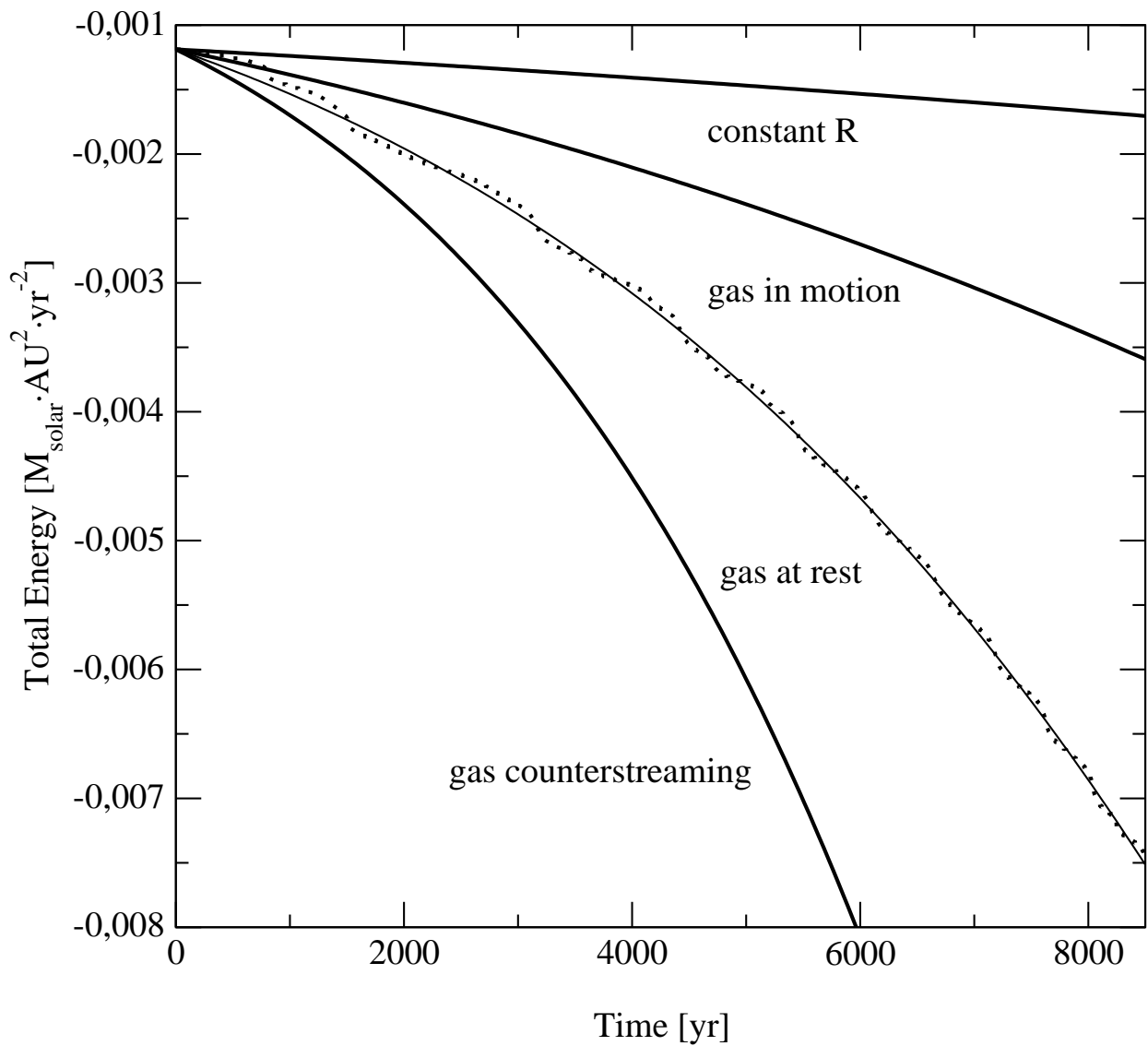}

\caption{\label{fig:Comparison-E(t)}Comparison of the numerical solution
(dotted line) of an accreting triple system accreting gas at rest
with the analytic solution. In addition the analytical solution for
accretion of gas in motion and of counter-streaming gas, as well as
the solution using the approximation $R=const.$ \citep{2001AJ....122..432R}
are shown. It can be clearly seen that the latter approximation underestimates
the absolute value of the total energy by an order of magnitude.}
\end{figure}

Fig. \ref{fig:Comparison-E(t)} illustrates the dependence of the total
energy over time for the different accretion models as well as the
$R=const.$-approximation, using the parameters of the Toy-Model in
\citet{2001AJ....122..432R}. Here it can be clearly seen that in the
$R=const.$-approximation the total energy differs by almost one order
of magnitude from our results, regarding accretion of gas at rest. 

To show the validity of our calculations, we also integrate the
equations of motion for a triple system accreting gas at rest
numerically as will be described in section \ref{sec:Simulations} and
plot the total energy of that system over time (dotted curve in
Fig. \ref{fig:Comparison-E(t)}). As it can be clearly seen, our
numerical result matches the predicted energy curve very well apart
from small deviations. These deviations, however, are not numerical
artifacts but caused by the assumption of virial equilibrium in our
analytical derivation, which is only strictly valid for a few special
cases of gravitationally interacting systems. The oscillations of our
numerical solution around the equilibrium one are compatible with the
oscillations of the virial coefficient $k$, defined as
$k=E_{kin}/\left|E_{pot}\right|$, around its equilibrium value of
$1/2$ in the study of
\citet{1989Ap&SS.158...19A} of gravitationally interacting,
non-accreting triple systems.

\begin{figure}
\plotone{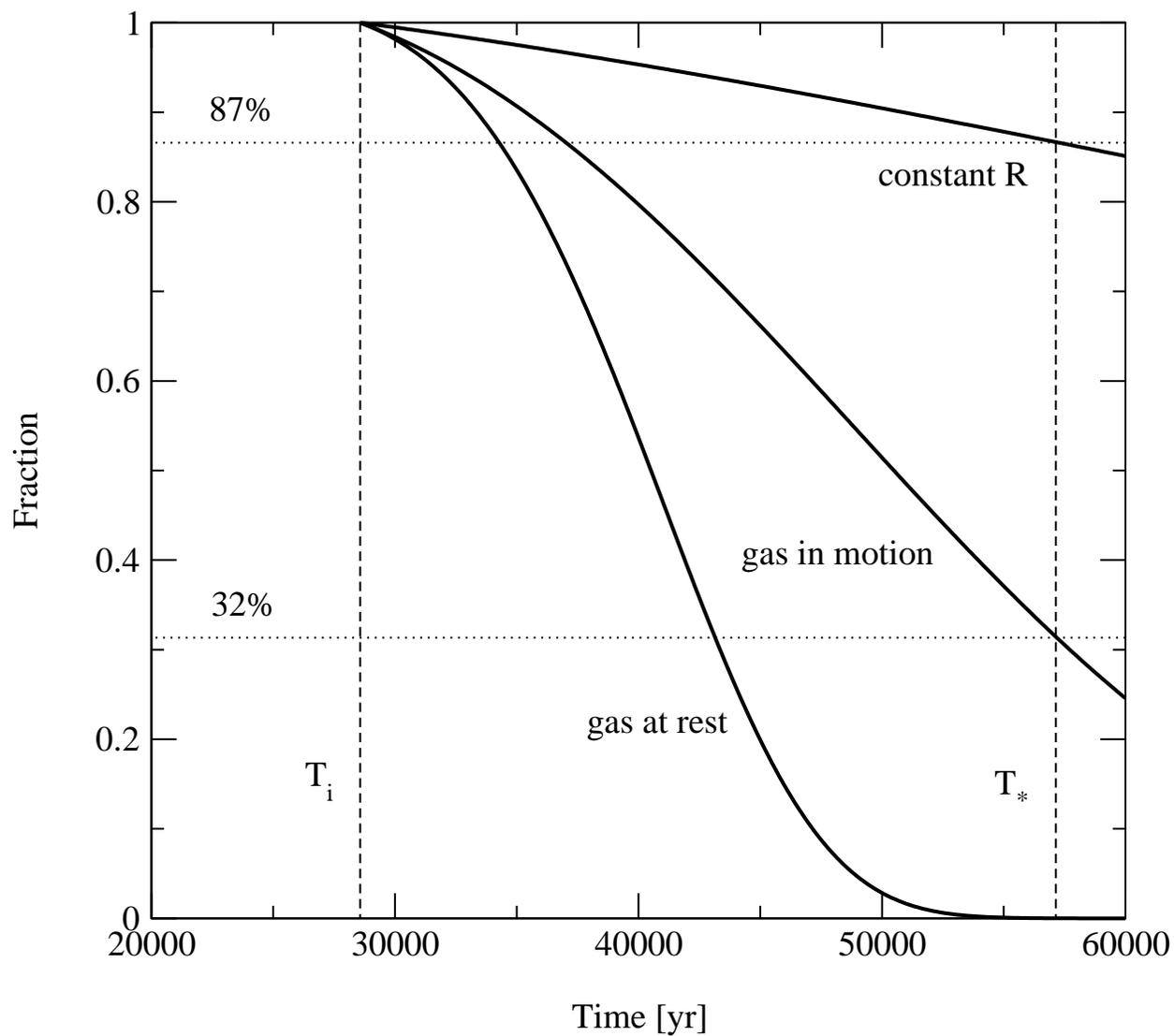}
\caption{\label{fig:decay-probability}The probability that an equal mass
triple system has not yet decayed after a time $t$ for the different
models. $T_{*}$ is the time when the fragments reach the Brown Dwarf
limit of $M=0.08{\rm M_{\sun}}$ and $T_{i}$ is the time the fragments
effectively start to interact with each other, which was chosen to
be the time when they reach $0.04{\rm M_{\sun}}$.}
\end{figure}

Since we found that both our analytically and numerically obtained
$E(t)$ agree very well, we now want to use the analytical solution
to obtain the formation probability of Brown Dwarfs repeating the
steps done by \citet{2001AJ....122..432R}, but using the half
life time of the decay for triple systems of $\tau=87t_{cr}$ as obtained
from our calculations. The result is shown in Fig. \ref{fig:decay-probability}.
 As it was to be expected from the energy curves of the previous section,
the $R=const.$-approximation underestimates significantly the number
of Brown Dwarfs. The approximation of 'gas in motion' leads to a six
times higher number of Brown Dwarfs and with the 'gas at rest'-approximation
almost all ejected fragments should be Brown Dwarfs. For our numerical
investigation that follows, however, we do not expect that all ejected
fragments have a mass lower than $0.08{\rm M_{\sun}}$. We find that
there is a significant number of meta-stable and stable systems with
extremely long decay times, challenging the assumption that the number
of systems that have not yet decayed is a simple exponential function
over time \citep{1987gady.book.....B}. Furthermore, the determination
of the half life of the decay is not unique, because counting the
number of systems that have not yet decayed after any other time as
$100t_{cr}$ gives very diverse values of $\tau$, sometimes deviating
by a factor of more than $2$. Clearly the question of how to obtain
statistically the number of decayed systems as a function of time
needs to be investigated more thoroughly. For this reason the decay
curves as shown in Fig. \ref{fig:decay-probability} are expected
to deviate quite a bit from our numerical results, but we do not expect
this deviation to be an order of magnitude.

\section{\label{sec:Simulations}Simulations}

\subsection{Initial Conditions}

In order to investigate the ejection scenario numerically we integrate
a large number (1000) of realizations of triple systems with an initial
mass of $0.04{\rm M_{\sun}}$ and with constant mass growth $\dot{M}$.
To cover all geometrically possible initial configurations we follow
the approach of \citet[Fig. 1]{1986Ap&SS.124..217A} where all three
bodies are initially in the $x-y$-plane and two bodies are always
placed at $x=-0.5,\, y=0$ and $x=0.5,\, y=0$. The position of the
third body is randomly chosen within a region lying in the positive
quadrant and embraced by a unit circle around the point $x=-0.5,\, y=0$.
This arrangement of the bodies has been proven to be a representative
sample for statistical studies of unstable triple systems by \citet{1994CeMDA..59..327A}.
We then multiply the initial position vectors by $200\,{\rm AU}$
to give the desired maximum separation.

The initial velocities of the cluster members are usually derived
from the kinematical properties of the surrounding molecular cloud
core. Since the observed ratio of rotational to potential energy of
molecular cloud cores, $\beta$, has been found to be low for most
of the cloud cores, with a value of $\beta=0.02$ \citep{1993ApJ...406..528G},
we do, for simplicity, neglect any possible initial uniform rotation
of our clusters. This decreases the lifetime of our triple systems,
but we do not expect this effect to be significant for our main results.
We also neglect any random motion the fragments could have due to
the thermal as well as the turbulent energy of the cloud. This approximation
seems justified as the temperatures of molecular cloud cores are rather
low and, because the turbulent velocities are thought to be subsonic
at the time fragmentation starts, the turbulent velocities must be
rather low as well. However, as we found from our simulations
most of our triple systems reached their virial equilibrium state
very quickly, after about two collisions, although we started from a
non-equilibrium state. Thus most of the observed collapse of the system is due to
its dissipative energy change rather than virialization, which is
why our results will not change significantly as a function of initial
velocities of the fragments. So, for simplicity, all our fragments
start with zero velocities getting the maximum value of ${\rm v}_{vir}$
for a given three body geometry. 

Our choice of initial positions and velocities results in a range of
total energies and consequently virial speeds with most of them having
values between $0.6$ and $0.8\,{\rm km\cdot s^{-1}}$. In this range
the distribution of ${\rm v}_{vir}$ is nearly flat. Thus our initial
conditions are, apart from the absolute value, very similar to the
ones favored by \citet{1998A&A...339...95S} and used by \citet{2003A&A...400.1031S}.

During the integration the fragments accrete mass at a given rate
which we will vary to investigate the influence of $\dot{M}$ on our
results. For $\dot{M}$ we choose 1, 2 and 5 times the value suggested
by \citet{2001AJ....122..432R} of
$\dot{M}_{RC}\approx1.4\cdot10^{-6}\,{\rm M_{\sun}\cdot yr^{-1}}$ per
fragment for a cloud with a temperature of $10\,{\rm K}$. We further
assume a certain radius around the origin outside of which the
accretion of the fragments is stopped if the system has decayed.  This
radius serves as an 'effective' cloud radius, determining the region
where the bodies accrete a significant amount of gas. Because we only
stop the accretion of a single body if the triple system has decayed,
we can investigate accreting equal mass systems and limit the
influence of the rather artificial geometry of the accretion region.
Otherwise the geometry of the accretion region would have strong
effects on our results through unequal accretion, which would not be
modelled reasonably in this case. In order to decide whether a system
has decayed, we employ two simple escape criteria. First, we require
that the escaper and the binary are unbound with respect to each
other, i.e. the total energy of the binary, treated as one body, and
the escaper is positive.  Second, we require that the distance between
the escaper and the center of mass of the binary is more than $7\cdot
d_{0}$ with $d_{0}$ being the initial mean harmonic distance. Hence
our cloud radius has to be larger than $7\cdot2/3\,
d_{0,max}\approx462\,{\rm AU}$ because as soon as the escaper reaches
the end of the accretion region we have to decide whether the system
has decayed and consequently whether we have to stop any further
accretion or not. The value of the minimum cloud radius was obtained
by taking into account that the escaper has twice the speed of the
binary and both are moving in opposite directions. Our cloud radius
was chosen to be $R_{cl}=600\,{\rm AU}$ and for comparison we also
performed some runs with higher $R_{cl}$ but found, that our results
do not vary significantly even if we increase this value by a factor
of 2.

As outlined in section \ref{sec:Analytical-Predictions}, we will investigate
two cases of momentum transfer during mass growth, one with zero momentum
transfer (gas is at rest) and one with a momentum that does not change
the velocity of the bodies, corresponding to gas that is accreted
while having the same direction and absolute value of the velocity
as the bodies. The latter case is there to unambiguously see the effect
when the bodies pick up momentum during accretion.

\subsection{Integration Method}

The integration of the equations of motion are performed with the
CHAIN code of \citet{1993CeMDA..57..439M}. This code gives sufficiently
low errors in the total energy budget. This is a necessary requirement
as the total energy directly relates to the ejection velocities, binary
separations and decay times we want to investigate. In order to deal
with accretion, the masses are updated after every CHAIN integration
step. When the masses are increased, the center of mass and velocity
as well as the momenta of the bodies are changed according to the
accretion model and the relevant reinitializations are done. In our
test calculations of constant-mass triple systems we observed a median
relative error in energy of $10^{-12}$.

\section{\label{sec:Results}Results}

\subsection{Number of Brown Dwarfs}

\begin{figure}
\plotone{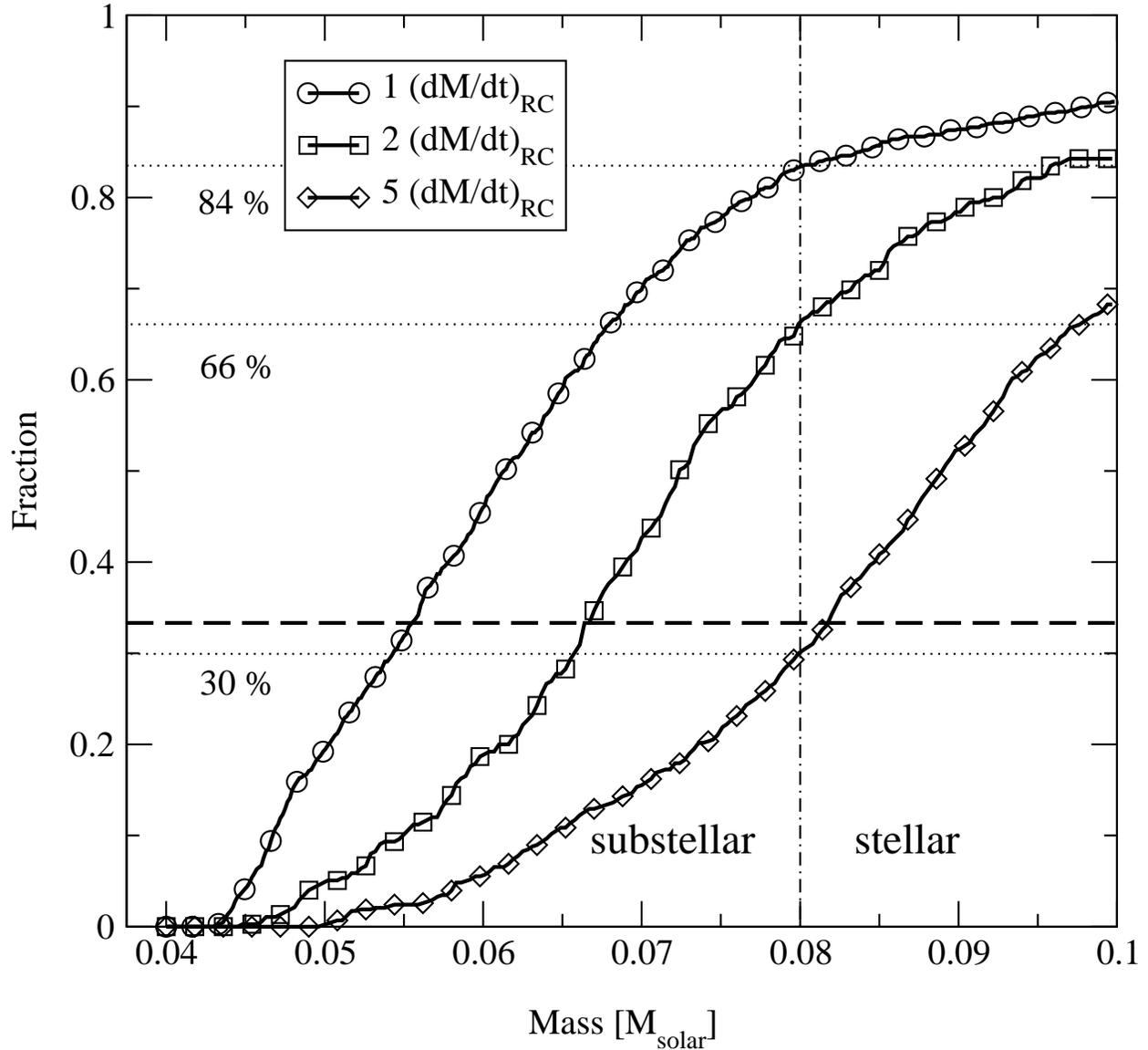}

\caption{\label{fig:BD number} Fraction of systems that ejected a single
member with a mass lower than a given mass $m$. Shown are the results
for different accretion rates in multiples of $1.4\,10^{-6}\,{\rm M_{\sun}yr^{-1}}$
of gas at rest. The dashed line represents the estimate of \citet{2001AJ....122..432R}
of ejected embryos with a lower mass than $0.08\,{\rm M_{\sun}}$.
It can be clearly seen that only if the accretion rate is 5 times
the value suggested by \citet{2001AJ....122..432R} the number of
Brown Dwarfs match their estimate.}
\end{figure}

In Fig. \ref{fig:BD number} %
the fraction of systems, accreting gas at rest, that ejected a single
member with a mass lower than a given mass $m$ is shown. As indicative
from the discussion in section \ref{sec:Analytical-Predictions} the
number of Brown Dwarfs in our simulation is lower ($84\%$) than the
analytically obtained value (nearly $100\%$). Since we measured the
mass of the ejected embryo at the cloud edge, the masses are generally
higher than at the time of decay which was predicted in our analytical
model. This difference, however, is only of the order of a few percent
and can, therefore, not explain the gap between our numerical and
analytical results. Furthermore, the influence of the accretion radius
on the decay times of the systems turned out to be
negligible. Nevertheless, the fraction of systems that produced Brown
Dwarfs is almost three times higher than was assumed by
\citet{2001AJ....122..432R} and more than six times higher as in the
$R=const.$-approximation for triple systems.  Considering that in
reality the accretion process is likely to be competitive, increasing
the formation probability of Brown Dwarfs, as outlined in section
\ref{sec:Analytical-Predictions}, our numerical results seem to
confirm that the ejection scenario can be very efficient even if one
only considers three fragments. It must be mentioned that the
{}``accretion-of-gas-at-rest approximation'' is only valid if the
fragments are moving at subsonic velocities
\citep{1997MNRAS.285..201B}. However, the average velocity of our
bodies is of the order of a few km$\cdot{\rm s^{-1}}$. Therefore we
still underestimate the decrease in energy of the systems, as
additional drag forces from the gas, caused by the bodies exciting
wakes in their passage, are not taken into account.

We also show in Fig. \ref{fig:BD number} the number of Brown Dwarfs
obtained at higher accretion rates. It can be seen that the number
of Brown Dwarfs decreases with increasing $\dot{M}$ and only if the
accretion rate is 5 times the value suggested by \citet{2001AJ....122..432R}
one gets about the same number of Brown Dwarfs as they obtained. This
once more demonstrates that the shrinkage of the system, reflected
by the time dependence of $R$, and the interaction with the gas decreases
greatly the total energy of the triple system at a given $\dot{M}$
and, therefore, increases the decay probability before the hydrogen-burning
limit is reached. It also shows that the decay curve over the time
$t$ cannot be expressed by a single exponential because of the significant
gap between our analytical and numerical results. 

Fig. %
\begin{figure}
\plotone{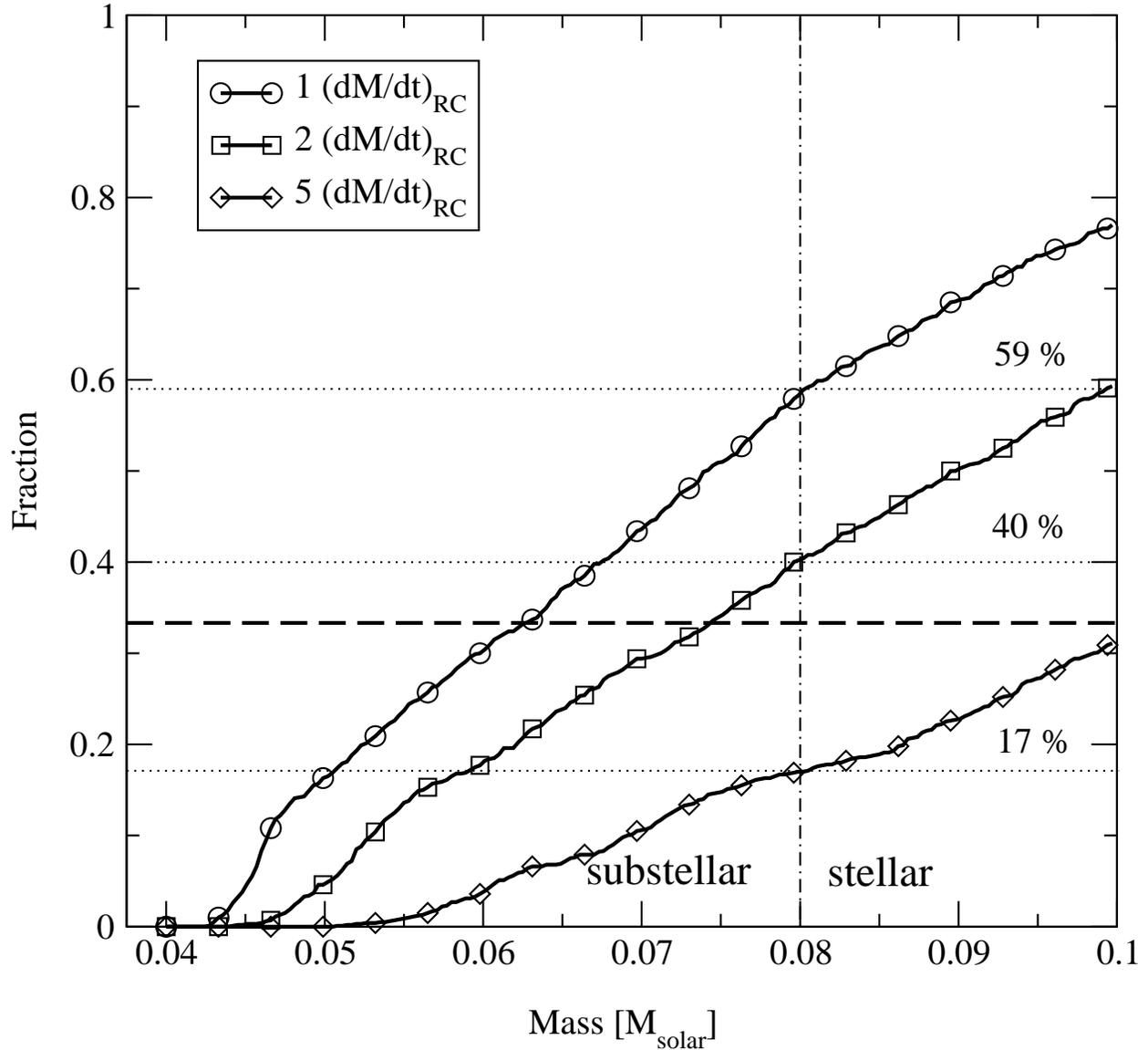}

\caption{\label{fig:BD number gas in motion} Fraction of systems that ejected
a single member with a mass lower than a given mass $m$. Shown are
the results for different rates of accretion of gas in 'extreme' motion
in multiples of $1.4\,10^{-6}\,{\rm M_{\sun}yr^{-1}}$. The dashed
line represents the estimate of \citet{2001AJ....122..432R} of ejected
embryos with a lower mass than $0.08\,{\rm M_{\sun}}$. As in the
case of accretion of gas at rest, the number of Brown Dwarfs in our
simulation is significantly higher than they assumed.}
\end{figure}
 \ref{fig:BD number gas in motion} shows the fraction of ejected
fragments with a lower mass than a given mass $m$, accreting gas
in 'extreme' motion. As in the case of accretion of gas at rest, the
number of Brown Dwarfs in our simulations is lower than we expected
from our analytical calculations. The difference, however, of about
9\% is considerably smaller than the difference in the case of accretion
of gas at rest, which was 18\%. Increasing the accretion rate causes
the number of Brown Dwarfs to decrease and in this case a little more
than twice the suggested value of $\dot{M}$ is necessary to obtain
the same number of Brown Dwarfs as anticipated by \citet{2001AJ....122..432R}.

\subsection{Decay Times}

\begin{figure}
\plotone{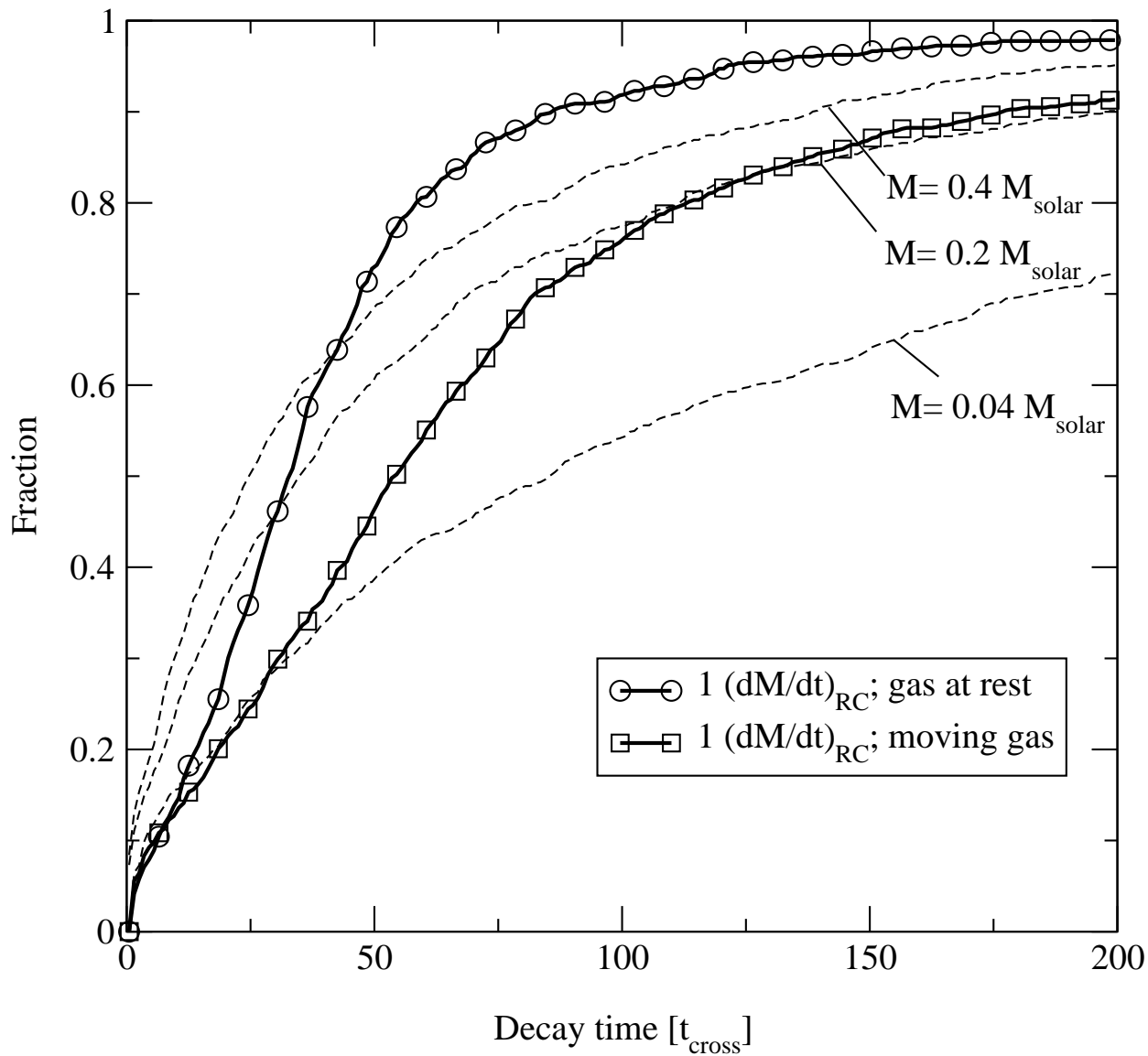}

\caption{\label{fig:decay times} Fraction of systems that decayed before
a time $t$ in initial ($M=M_{0}$) crossing times. The solid lines
represent the results for triples accreting gas at rest (open circles)
and for triples accreting gas in motion (open squares). For comparison
the results of decaying triple systems with constant mass of $M=0.04,\,0.2,\,0.4\,{\rm M_{\sun}}$
are also shown (dashed lines). }
\end{figure}

In Fig. \ref{fig:decay times} %
 the cumulative distribution of the decay times of accreting triple
systems is shown and compared to those of decaying systems of constant
mass. As it was expected due to the decrease in total energy over
time, the decay probability of accreting triples quickly exceeds the
time for a constant mass of $M=0.04\,{\rm M_{\sun}}$ and after some
crossing times even those with $M=0.2\,{\rm M_{\sun}}$ and $M=0.4\,{\rm M_{\sun}}$.
The different slopes of the curves for the different accretion models
reflect the different time dependence on the total energy. Due to
the different time dependence, the decay probability of the triple
systems accreting gas at rest is generally higher than that of the
triple systems accreting gas in motion. It is worth pointing out that
the maximum fragment mass reached in our experiments with accretion
is still lower than $0.4\,{\rm M_{\sun}}$ for the accretion of gas
at rest model and lower than $0.2\,{\rm M_{\sun}}$ in the case of
accretion of gas in motion. This might look counterintuitive because,
on average, lower-mass but accreting triple systems seem to decay
earlier than heavier non-accreting systems. However this is a direct
consequence of the time dependence of $R$ explained in section \ref{sec:Analytical-Predictions}.

\subsection{Escape Velocities}

Fig. %
\begin{figure}
\plottwo{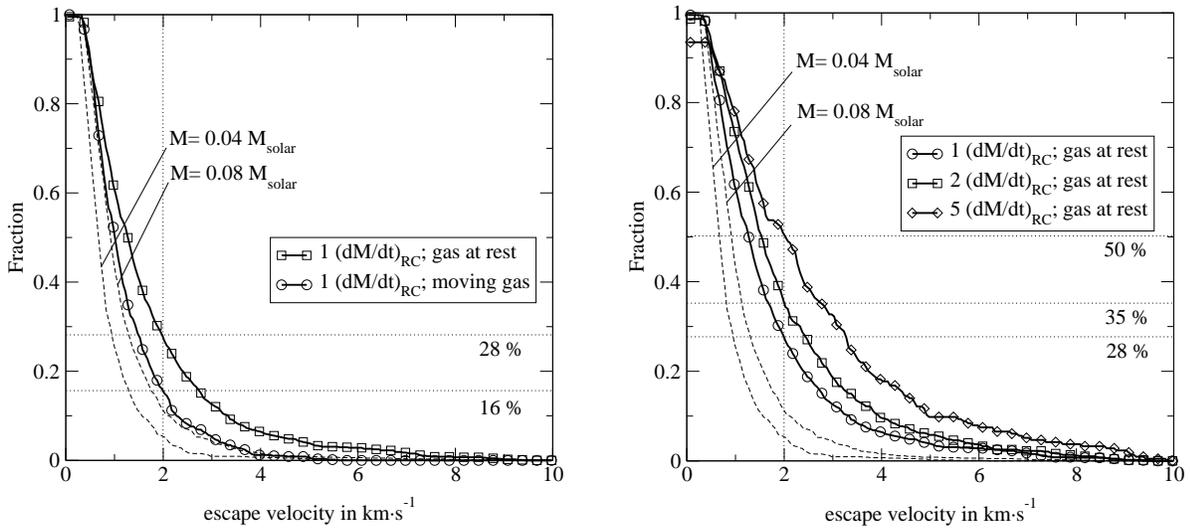}{f6b.eps}

\caption{\label{fig: escape-vel} Fraction of systems, producing a single
ejected Brown Dwarf with a speed larger than a given velocity v for
different accretion models (left) and different accretion rates (right).
Also shown are the results for systems of constant mass (dashed lines). }
\end{figure}
\ref{fig: escape-vel} shows the distribution of escape velocities
of ejected Brown Dwarfs which are higher than a certain velocity v.
Compared to constant-mass systems, the escape velocities are
considerably higher for triple systems accreting gas at rest with the
median of the distribution being boosted up by a factor of
two and more, depending on the accretion rate. 

In our simulations more than 28\% of the escaping Brown Dwarfs
have a larger velocity than $2{\rm km\cdot s^{-1}}$ as opposed to only
10\% in the case of constant-mass systems of $M=0.08$.  One also finds
that there is a tendency towards higher escape velocities with higher
accretion rates with more massive Brown Dwarfs being formed at higher
$\dot{M}$. Half of the Brown Dwarfs that formed at an accretion rate
of $5\dot{M}_{RC}$ obtain an escape velocity of more than $2\,{\rm
km\cdot s^{-1}}$.

For systems that accrete gas in 'extreme' motion,
the escape velocities are only marginally higher than the velocities
for systems of constant mass with $M=0.08\,{\rm M_{\sun}}$. The
difference to the escape velocities from systems accreting gas at rest
is a direct consequence of the, on average, lower absolute value of
the total energy at decay time of the systems accreting gas in
motion. In all cases the escape velocities of Brown Dwarfs are always
higher if they have been ejected from accreting triples even if
compared to the ones from the most massive equal-mass Brown Dwarf
system with constant mass.

These results have a direct implication for the abundance of Brown
Dwarfs in stellar systems as discussed by \citet{2003MNRAS.346..369K}.
As we find almost more than a third of the escapers having velocities
exceeding $2\,{\rm km\cdot s^{-1}}$, they should easily escape their
stellar birth cluster if this cluster has a lower escape velocity
than the Taurus cluster with ${\rm v}_{esc}=1.4\,{\rm km\cdot s^{-1}}$.
Therefore, it is observationally difficult to detect such free-floating
Brown Dwarfs at an advanced age like that of Taurus. On the other
hand, it must be emphasized that for accretion rates which are not
too high the fraction of escapers with a lower escape velocity than
$1.4\,{\rm km\cdot s^{-1}}$ is, with about 50\%, very high and one
should therefore also expect many Brown Dwarfs to be retained in similar
clusters at a younger age. The observed abundance of Brown Dwarfs
in a young cluster also depends critically on the evolution of the
whole cluster which is until today only well understood for a few
of the sufficiently young star-forming regions, like Taurus, the Pleiades
and the ONC. This makes it hard to draw any conclusions about how
Brown Dwarfs form from the observed abundance alone in other clusters,
where there is only little information about the earlier cluster evolution.

\subsection{Binary Semi-Major Axis}

The investigation of binary properties should give tighter constraints on
the formation model as at least for close binaries it is not expected that
they change their orbital parameters a lot due to possible interactions
with other members of the surrounding stellar cluster. If the surrounding
cluster is virialized, the encounter probability for sufficiently close
encounters is only significant in regions with extreme stellar densities,
like the inner $0.4\,{\rm pc}$ of the Trapezium cluster (compare to
\citealp{1991MNRAS.249..584C}).

The advantage of our simulations is the ability to investigate a large
number of Brown Dwarf binaries, due to the neglect of competitive
accretion. As in reality, however, competetive accretion will take place to
some degree, we investigated this effect by performing test calculations of
accreting triple systems, where the same infalling mass is unequally
distributed. We found a clear trend towards lower abundances of Brown Dwarf
binaries with higher differences in the accretion rates, while the
semi-major axis distribution did not change significantly. From the latter
result and the assumption that the two heaviest bodies always form the
binary, the reason for the lower abundance is simply a result of the
binaries having higher masses for a given total energy at the time of decay
which results in lower escape velocities. 

In all of our runs, we got a few hundred up to more than 800 binary
Brown Dwarfs, depending on the accretion rate, which makes it possible
to obtain statistically meaningful results. In Fig. %
\begin{figure}
\plotone{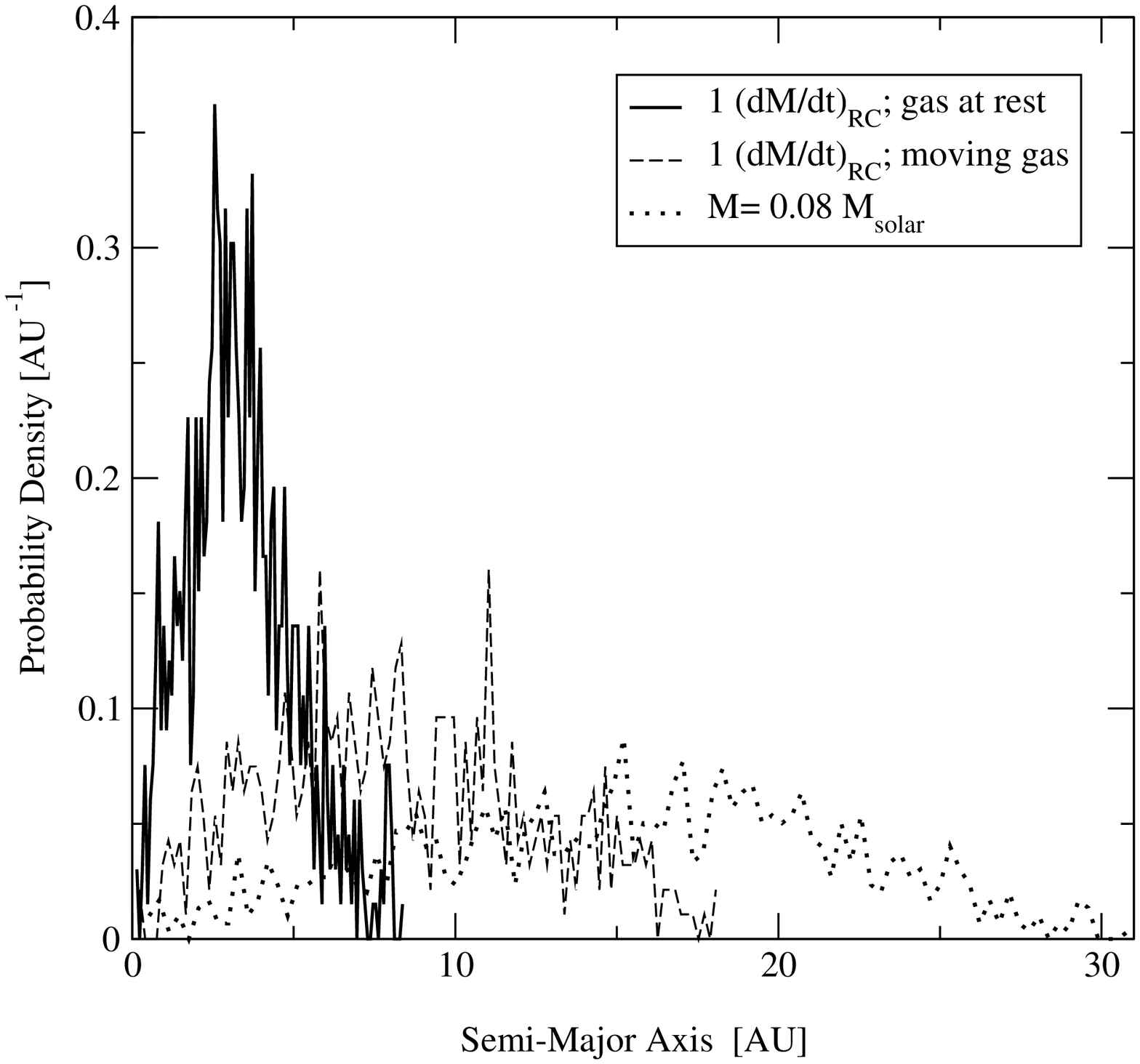}

\caption{\label{fig:semi-major} Semi-major axis distribution for different
kinds of accretion at $\dot{M}=\dot{M}_{RC}$ as well as for constant
mass systems with $M=0.08\,{\rm M_{\sun}}$. Due to the steeper decrease
of the total energy in the case of accretion of gas at rest the resulting
Brown Dwarf binaries have lower separations than in the case of accretion
of gas in 'extreme' motion. }
\end{figure}
\ref{fig:semi-major} the distribution of the semi-major axis of our
obtained Brown Dwarf binaries is shown for the different types of
accretion as well as for $M=0.08\,{\rm M_{\sun}}$-systems of constant
mass. In this plot it is remarkable that the separations of the
heaviest possible Brown Dwarf binaries resulting from the decay of
systems with constant mass are on average a lot higher, with its
median being larger by a factor of 5, than the binary separations
resulting from accreting lower mass triples. It can be clearly seen
that the semi-major axis distribution for accreting triple systems is
narrower and the peak is at lower values compared to systems of
constant mass. The difference is again caused by the time dependence
of the total energy for the different types of accretion compared to
the $R=const.$ case. It is also responsible for reducing the median
binary separation to 1/50th of its original value of $R=200$. Also
from this plot one can infer that the more the momentum of the bodies
is reduced the narrower the semi-major axis distribution and the lower
the peak position of the distribution will get. We also found that the
peak of the semi-major axis distribution does not change much with the
accretion rate, only the cut-off is lower for higher rates. This is
simply a result of only considering Brown Dwarf systems, because the
final total energy is $\sim(M(t)/M_{0})^{n}$, with $n$ depending on
the accretion model, which means that the total energy depends on the
final mass of the fragments and not directly on the accretion
rate. Therefore choosing a maximum mass limit determines the available
maximum energy that can be divided between the kinetic energy of the
escaper and the energy of the binary. The lower cut-off of the
distribution for higher accretion rates is mainly due to the fact
that, at higher accretion rates, more massive Brown Dwarf binaries are
formed, which causes the minimum total energy of the system to
increase thus decreasing the maximum semi-major axis.

Surprisingly the semi-major axis distributions resulting from accreting
triple systems, bear already some resemblance to the observed one
by \citet{2003AJ....126.1526B}, as there is no Brown Dwarf binary
with a wider separation than $20\,{\rm AU}$ and the observed peak
is at about the same value as the observed value $a_{peak}\approx3\,{\rm AU}$.
The observed sample, however, has the disadvantage that it is magnitude
limited and therefore prone to biases. 

\begin{figure}
\plotone{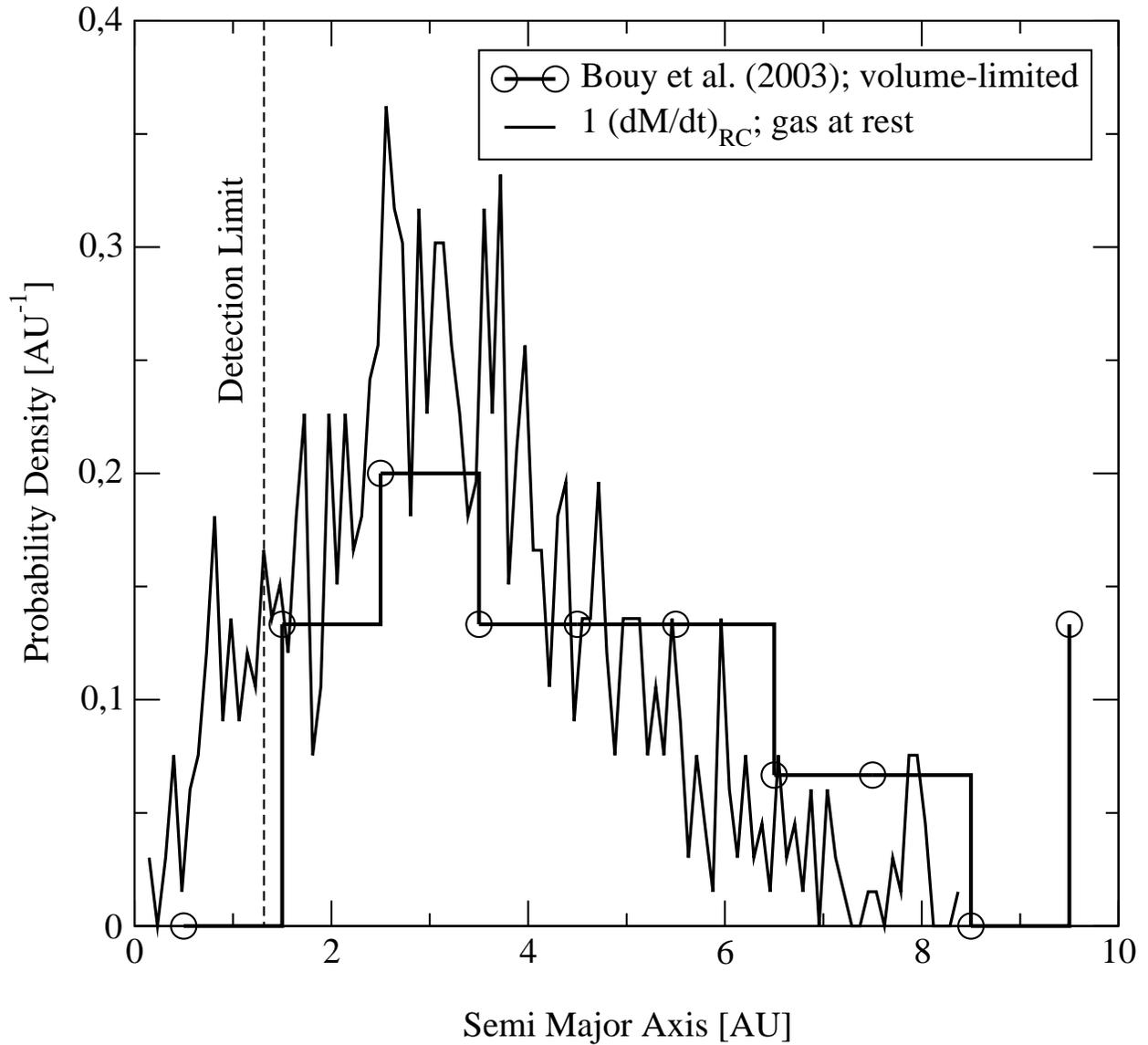}

\caption{\label{fig:semiMajorBouy} Semi-major axis distribution of the Brown
Dwarf binaries obtained in our simulation of decaying triple systems,
accreting gas at rest, and the observed volume-limited sample distribution
of \citet{2003AJ....126.1526B}. These two distribution match very
well, given the uncertainties (Poisson noise) which are of the order
of about a third for the observed distribution. Both distributions
have a peak at about the same value of $a=3\,{\rm AU}$ and show about
the same degree of asymmetry around the peak.}
\end{figure}

To get an approximately unbiased sample it is better to choose the
binary Brown Dwarfs over a finite volume within which both components
can be fully resolved. For the sample of \citet{2003AJ....126.1526B}
this would include all binaries with a distance of less than $25\,{\rm pc}$
(Brandner, private communication). The semi-major axis distribution
of these objects is shown in Fig. \ref{fig:semiMajorBouy}%
together with our numerically obtained distribution for decaying
triples, accreting gas at rest. There is a remarkable agreement
between these two distributions as they both have the peak value at
$a\approx3\,{\rm AU}$ and approximately the same, rather steep, slope
to both sides of it. That our distribution fits almost perfectly the
one obtained from observations is certainly a coincidence given the
arbitrariness of our initial conditions. However, what we can predict
is that, according to our model, the distribution cannot be flat for a
semi-major axis below the peak value, but must decrease. Due to the
detection limit of the observations of \citet{2003AJ....126.1526B} it
is not clear whether the observed distribution will also decrease with
decreasing $a$ or not. This rather steep decline in our simulation is
not a numerical artifact but rather seems to be a general property of
binaries formed directly by decaying multiple systems without
dissipative processes, such as tidal interactions between the
fragments. These effects are known to form much harder,
spectroscopic binaries resulting in a flatter semi-major axis
distribution for small separations like it was observed for G-stars by
\citet{1991A&A...248..485D}. If observationally confirmed, 
the decrease of the semi-major axis distribution to lower separations
would indicate that tidal interactions are less important for the
formation of Brown Dwarf binaries.

The asymmetric shape of our semi-major axis distribution is similar
to the one in \citet[Fig. 9b]{1998A&A...339...95S}, which was obtained
by integrating constant mass triple systems with initial masses constrained
by an observed stellar IMF as well as a clump mass spectrum for the
total cluster mass and scaling the results assuming a constant virial
speed. The interesting point here is that our decaying accreting triple
systems with equal initial masses seem to \emph{result} in \emph{}a
very similar total energy distribution after they stopped accretion
without applying any constraints. It should be mentioned that our
semi-major axis distribution is not strongly influenced by the deviation
of our initial virial velocities from the ${\rm v}_{vir}=const.$-assumption
of \citet{2003A&A...400.1031S}. We found that even if we choose only
triple systems which have their initial ${\rm v}_{vir}$ in any very
narrow range within the flat part of our initial virial velocity distribution,
the asymmetric shape of the semi-major axis distribution is recovered
and, moreover, the obtained distributions hardly differ from the one
shown in Fig. \ref{fig:semiMajorBouy}.

\section{\label{sec:Conclusions}Conclusions}

In this work we investigated the influence of accretion and gas
interaction on the decay of gravitating triple systems and its
implications on the viability of the ejection scenario as Brown Dwarf
formation scenario. We have shown that accretion and consequent
shrinkage of triple systems increases the velocities significantly
compared to non-accreting systems. The high velocities of Brown Dwarfs
should make it less likely to find them in star-forming clusters with
a shallow potential well and a low escape velocity, but rather in an
extended region around it, usually called the Brown Dwarf halo. This
would, at a first glance, contradict the rather high abundance of
Brown Dwarfs observed in Chamaeleon which have the same spatial
distribution as the stellar population \citep{2004A&A...416..555L},
but one must also bear in mind that there is still a substantial
fraction of Brown Dwarfs in our simulations having escape velocities
which are rather low compared to typical escape velocities of low-mass
star-forming clusters. These Brown Dwarfs should also share the same
kinematics as the stars in the cluster, which would explain that the
velocity dispersion of Brown Dwarfs in Chamaeleon is very similar to
the one of the stars \citep{2001A&A...379L...9J}. Given that the
formation probability of Brown Dwarfs in the ejection scenario can be,
in principle, rather high, the high abundance of them in Chamaeleon
does not seem unreasonable as long as there is no reliable estimate of
the combined potential of stars and the remaining gas. Since in this
region there are only a few massive stars, gas removal should be less
efficient, therefore increasing the influence of the gas
potential. The abundance and spatial distribution of Brown Dwarfs
within star-forming regions seems to depend on many parameters, which
are observationally not easily accessible making it rather hard to
argue convincingly for or against a Brown Dwarf formation model based
on these criteria.

Tighter constraints on a Brown Dwarf formation model should be expected
from Brown Dwarf binary properties. As already mentioned in section
\ref{sec:Introduction} the Brown Dwarf binary properties observed so far do
not seem to be a natural extension of the binary properties obtained from
the standard star-formation model of \citet{2003MNRAS.346..369K} for
hydrogen-burning stars. In general the components seem to be very close and
their semi-major axes are distributed in a rather narrow region below
$\approx20\,{\rm AU}$. As our simulations show such features are readily
obtained if accretion is taken into account during decay. The median
of the binary semi-major axis distribution in the case with accretion
turned out to be up to 50 times smaller compared to the initial
inter-particle distances.  This reduction in scale made it possible to
start with much larger fragment separations of a few $100 \rm AU$ in order
to obtain the observed close Brown Dwarf binaries. This is also the typical
length scale on which fragments are formed in collapse calculations of
molecular cloud cores. To achieve the same Brown Dwarf binary properties
without accretion the fragments must be brought initially in almost
unreasonably close configurations as explained in section
\ref{sec:Previous-Work}, because the typical binary separation is by a
factor of up to 5 larger than in the case with accretion.

We were also able to represent the semi-major axis
distribution of the volume limited sample of binary Brown Dwarfs of
\citet{2003AJ....126.1526B}. A general feature of this distribution is
a rather asymmetric shape which can also be obtained by decaying
constant mass systems with masses constrained by observed clump and
stellar mass spectra as well as assuming a constant virial speed
\citep{1998A&A...339...95S,2003A&A...400.1031S}.  The important point
to make here is that we were able to \emph{produce} such a semi-major
axis distribution without any mass constraints other than our initial
masses. Another feature of these distributions is a steep drop off
to both sides of the median, which is not observed for G-type binaries
\citep{1991A&A...248..485D}. This is mainly because we did not include
such dissipative processes like tidal interaction between fragments as
well as interactions between their disks, which tend to reduce the
binary separations further and circularize their orbits. On the other
hand, if this drop off to lower separations is confirmed
observationally, it would provide some evidence, that tidal
interactions might be less important for the evolution of Brown Dwarf
binaries than they are for G-type stars. In general the relative
numbers of spectroscopic binaries and those near the median separation
should provide evidence for or against a dynamical origin of an
observed binary distribution. We also found that the Brown Dwarf binary
distribution is not much affected by competitive accretion, as our
test calculations of unequally accreting triple systems indicate. This
would mean that the distribution depends only to a lesser degree on
the detailed distribution of mass within the triple system, but needs
to be confirmed in future studies.

Given the similarities in our
distribution to \citet{1998A&A...339...95S},it is rather hard to
judge from observations how large the influence may be of other
constraints like the maximum total mass of the cluster determined by
the mass of the molecular cloud core.

To answer this question it would be necessary to investigate how the
initial properties of forming clusters relate to the properties of
their parental molecular cloud core. Recent studies seem to indicate
that they depend strongly on the remaining turbulent velocity field
\citep{2004A&A...414..633G,2004MNRAS.347..759D}, but there is yet
no detailed investigation on how this influences the initial properties
of the clusters. \citet{2004MNRAS.347..759D} find wider configurations
in their simulations when the index of the turbulent power spectrum
$\alpha$ is as high as $-5$, and closer ones for lower indices,
but do not report on other quantities like initial cluster energy
or virial state of the cluster. They further find that for an index
of $-5$ there are fewer Brown Dwarfs forming than at an index of
$-3$. They explain this fact by the occurrence of wider configurations
at the $\alpha=-5$-case. An alternative explanation could be given
by our model of decaying triple systems accreting gas in 'extreme'
motion, as in the case of $\alpha=-5$ the turbulent motion of the
gas is on larger scales and, therefore, it is more likely that the
accreted gas is adding some momentum to the bodies, even though not
as high as we did in our simulations.

We have shown analytically as well as numerically that the probability
of forming Brown Dwarfs should be high even for initially moderately
compact systems. This is also true without considering
competitive accretion, as competitive accretion will increase the
number of Brown Dwarfs, which also our test simulations of
unequally accreting triple systems indicate. Only if the accretion
rate is very high, in our simulations of the order of $10^{-5}\,{\rm
M_{\sun}\cdot yr^{-1}}$, forming Brown Dwarfs by ejection seems less
likely especially if the accreted gas changes the momentum of the
bodies. Our analytical calculation showed furthermore that the reason
why the ejection scenario is much more efficient than previously
assumed lies in the shrinkage of the system, reflected by the time
dependence of $R$, which causes the energy to decrease further
(compare to eqn. \ref{eq:Edot(v)}). It turned out that the total
energy, assuming accretion of gas at rest, is proportional to
$(\dot{M}/M_{0}\cdot t)^{5}$ and, assuming gas in extreme motion, it
is proportional to $(\dot{M}/M_{0}\cdot t)^{3}$ while the time
dependence under the $R=const.$-approximation is only
$\sim(\dot{M}\cdot t)^{2}/R$.  This convincingly explains the very
different formation probabilities, resulting from our numerical
calculations for the different kinds of momentum transport during mass
growth, even though our analytical results differ significantly from
our numerical ones. The differences must be due to the fact that the
assumption, that the time of the decay can be described as a single
exponential function with a half life directly proportional to the
crossing time, is not strictly valid, as even varying the half life of
the decay did not reduce this difference significantly.

We conclude that accretion of gas and the kinematic properties of
the accreted gas during dynamical interactions strongly influence
the abundance as well as the dynamical properties of Brown Dwarfs
and because of the high formation probability and the agreement between
our semi-major axis distribution and the observed one of \citet{2003AJ....126.1526B},
makes the ejection scenario a viable option for forming single as
well as binary Brown Dwarfs if only three fragments are involved.

\acknowledgements{We like to thank Wolfgang Brandner for his support and advice on
obtaining the volume limited sample of the observed binary Brown Dwarfs
of \citet{2003AJ....126.1526B}.}

\appendix

\section{\label{sec:Energy-analytic}The Energy of Accreting Multiples}

To obtain an expression for the energy of accreting bodies we first
have to consider how much momentum the infalling gas carries onto
them. The first case we want to look at is accretion of gas at rest,
which means that the momentum of the fragments is not changed when
the mass is increased. The change of the momentum is therefore only
due to the gravitational force between the bodies. The Newtonian equations
therefore read
\begin{equation}
\frac{dM_{i}}{dt}\cdot{\bf v_{i}}+\frac{d{\bf v_{i}}}{dt}\cdot
M_{i}=G\,{\displaystyle \sum_{i\neq
j}}\left(\frac{M_{i}M_{j}}{r_{ij}^{2}}\frac{{\bf
r_{ij}}}{r_{ij}}\right).\label{eq:Egasatrest}
\end{equation}
To derive the energy equation for multiple systems, accreting gas
at rest, we multiply both sides of equation \ref{eq:Egasatrest} with
${\bf v_{i}}$ and rearrange it to give \begin{equation}
\frac{dE}{dt}=-{\displaystyle \sum_{i}}\left(\frac{dM_{i}}{dt}\frac{v_{i}^{2}}{2}\right)-G\,{\displaystyle \sum_{i\neq j}}\left(\frac{d(M_{i}M_{j})/dt}{r_{ij}}\right)\label{eq:Edot(v)}\end{equation}
with\begin{equation}
E={\displaystyle \sum_{i}}\left(\frac{M_{i}}{2}v_{i}^{2}\right)-G\,{\displaystyle \sum_{i\neq j}}\left(\frac{M_{i}M_{j}}{r_{ij}}\right).\end{equation}
Splitting up $E$ in kinetic and potential Energy, $E_{kin}$ and
$E_{pot}$, equation \ref{eq:Edot(v)} becomes\begin{equation}
\frac{dE}{dt}=-\frac{\dot{M}}{M}E_{kin}+2\,\frac{\dot{M}}{M}E_{pot}\end{equation}
where we set $M_{i}=M$ and $\dot{M}_{i}=\dot{M}$ for all $i$ as
we will only consider equal mass systems in this paper. Assuming virial
equilibrium, that gives \begin{equation}
\frac{1}{2}E_{pot}=E\end{equation}
and\begin{equation}
E_{kin}=-E\,,\end{equation}
finally leads to \begin{equation}
\frac{dE}{dt}=5\cdot\frac{\dot{M}(t)}{M(t)}\, E.\label{eq:Edot(E)rest}\end{equation}
This is a linear first-order differential equation with time varying
parameters and has the solution\begin{equation}
E(t)=E_{0}\cdot\exp\left({\displaystyle \int\limits _{0}^{t}}5\cdot\frac{\dot{M}(u)}{M(u)}\, du\right).\end{equation}
By setting $M=\dot{M}\cdot t+M_{0}$ and $\dot{M}=const.$ we obtain
the final energy equation for our model with accretion of gas at rest\begin{equation}
E(t)=E_{0}\cdot\left(\frac{\dot{M}}{M_{0}}t+1\right)^{5}.\end{equation}

To see how the energy decreases if the gas is not at rest, we assume,
for simplicity, that the gas is always moving in the same direction
and at the same speed as the accreting body. The equations of motion
remain unchanged in this case and the energy relation reads
\begin{equation}
\frac{dE}{dt}=+\frac{\dot{M}}{M}E_{kin}+2\,\frac{\dot{M}}{M}E_{pot}
\end{equation}
and repeating the steps we did to obtain $E(t)$ for the accretion
of gas at rest, we finally get
\begin{equation}
E(t)=E_{0}\cdot\left(\frac{\dot{M}}{M_{0}}\cdot t+1\right)^{3}
\end{equation}
which differs from the solution with accretion of gas at rest only
in the value of the exponent.

\bibliographystyle{apj}
\bibliography{BDFormationReport}

\end{document}